\documentclass[usegraphicx,usenatbib]{mn2e}
\usepackage{times}

\def\apj {ApJ}

\def\apjs {ApJS}
\def\aj {AJ}

\def\mnras {MNRAS}
\begin{document}
\title[Properties of galaxies in groups]
{Groups of galaxies: relationship between environment and galaxy properties} 
\author[H.J. Mart\'{\i}nez \& H. Muriel]
{H\'ector J. Mart\'{\i}nez\thanks{E-mail: julian@oac.uncor.edu}
\& Hern\'an Muriel\\
Grupo de Investigaciones en Astronom\'{\i}a Te\'orica y Experimental,
IATE, Observatorio Astron\'omico, Universidad Nacional de C\'ordoba,\\
Laprida 854, X5000BGR, C\'ordoba, Argentina \\
Consejo de Investigaciones Cient\'{\i}ficas y T\'ecnicas (CONICET),
Avenida Rivadavia 1917, C1033AAJ, Buenos Aires, Argentina\\
}
\date{\today}
\pagerange{\pageref{firstpage}--\pageref{lastpage}} 
\maketitle
\label{firstpage}
\begin{abstract}
We analyse how the properties of galaxies in groups identified
in the Sloan Digital Sky Survey depend on the environment.
In particular, we study the relationship between galaxy properties and
group mass and group-centric distance.  Among the galaxy properties
we have considered here, we find that the $g-r$ colour is the
most predictive parameter for group mass, while the most predictive pair of properties
are $g-r$ colour and $r-$band absolute magnitude.
Regarding the position inside the systems, the $g-r$ colour is the best tracer
of group-centric distance and the most predictive pair of properties are $g-r$ colour
and spectral type taken together.
These results remain unchanged when a subsample of high mass groups is analysed. The 
same happens if the brightest group galaxies are excluded.
\end{abstract}
\section{Introduction} 
It is well known that galaxy properties correlate with the environment, e.g.,
the morphology-density relation \citep{dress80}, star-formation-density relation 
\citep{gomez}. Nevertheless, the detailed joint distribution of these properties 
as a function of the galaxy clustering remains unclear. 
As all the galaxy properties (e.g. morphology, luminosity, colour) are correlated
it is not surprising that all properties correlate with environment, but
the question that arises is which of the properties are correlated to environment
independently of the others.
In a recent paper, \citet{blanton05} systematically explore the local environment
of galaxies in the Sloan Digital Sky Survey \citep[SDSS;][]{york00}, as a 
function of their luminosity, surface brightness, colour and S\'ersic index.
These authors find that colour is the galaxy property most predictive 
of the local environment for field galaxies. 
They also analyse pairs of properties taken together, finding that
galaxy colour and luminosity jointly comprise the most predictive one.

While clusters of galaxies have been intensively studied over the last
decades, detailed studies of galaxy groups and their evolution have only
recently begun.   
The study of the properties of galaxies in intermediate mass systems
is particularly important to understand how galaxies evolve and how different 
physical mechanisms affect them. In particular, galaxy interactions are expected
to be more common in groups than in clusters. In groups, velocity dispersions
are typically not much larger than that of the member galaxies.
It has been argued that the high fraction
of early-type galaxies in clusters is mainly the result of galaxy-galaxy 
interactions within groups \citep{zm98,zm00,ho00}.

\citet{yo02} showed how the fraction of star forming galaxies goes down
as a function of group mass. \citet{mardom02} found a similar trend 
with decreasing  group centric distance. 
It is well known that there are correlations between different galaxy
properties. Therefore, the observed trend of increasing fraction of non star
forming galaxies (a spectroscopic property) with group mass, implies a similar 
relation between the fraction of red galaxies (a photometric property) or the
fraction of bulge-dominated galaxies (a morphological property) with group mass.
In this work we address the following questions: `which galaxy properties
are more tightly correlated with group mass?' and `which
galaxy properties are more affected by the position inside a group?'.
This paper is organised as follows: in section 2 we describe the sample of 
galaxies in groups we use; while the analysis of the dependence of galaxy properties
on mass and on the group-centric distance are carried out in sections 3 and 4
respectively. We summarise our results and discuss them in section 5.
\section{The sample of galaxies in groups}
The sample of galaxies in groups used in this paper has been taken from the
group catalogue identified by \citet{zmm06}.
This catalogue was constructed from the Main Galaxy Sample \citep[MGS;][]{mgs}
of the Fourth Data Release of the SDSS \citep[DR4;][]{dr4}
following the same procedure as in \citet{mz05}. It consists
of a standard friend-of-friend algorithm for group identification, the 
application of a procedure to avoid artificial merging of smaller systems in
high density regions and an iterative method to compute reliable group 
centre positions. The catalogue, in its improved identification version, 
includes 14004 galaxy groups with at least 4 members in the area spectroscopically
surveyed by DR4, accounting for a total of 85687 galaxies. 

We have chosen to work with volume limited samples of galaxies instead of using flux 
limited ones and individual galaxy weights according to their luminosities.
We find that the results are more robust in a volume limited sample. In a magnitude
limited sample, the galaxy population observed in the nearest groups is significantly
different than that observed for the most distant ones. Something similar happens
with the groups' parameters. Therefore, observational effects such as seeing that
affect certain galaxy parameters can introduce systematic effects in our statistics.
In section 3 we present some tests in order to evaluate the reliability of the 
results. We have restricted our analysis to galaxies in groups 
down to $M_r-5\log(h)=-18$ and up to a conservative maximum redshift 
of $z_{\rm max}=0.043$, 
that gives a volume limited sample according to the selection criteria of the MGS.
Another relevant reason to choose a small value for $z_{\rm max}$ is related
to two of the galaxy parameters we are considering in our analysis: the concentration
parameter, defined as the ratio of the radii that enclose 90\% and 50\% of the 
Petrosian flux and the surface brightness, that involves the latter. 
Seeing affects the determination of those radii, and this becomes
more important for more distant galaxies that have smaller angular size. 
Our sample consists of 6183 ($M_r-5\log(h)\leq-18$) galaxies in 1691 groups.
From this sample we construct a number of subsamples of galaxies defined by group
virial mass and number of members, that are detailed in the analysis sections. 
Throughout this work, we use the virial mass and virial radius of groups 
computed by \citet{zmm06}.  Therefore, when we refer to `group mass',
it should be remembered that we are dealing with `group virial mass'.
The mass distribution for our sample of groups is shown in the lower right
panel of Figure \ref{fig1}.

\begin{table}
\caption{Adopted parameter cut-offs}
\begin{tabular}{lrr}
\hline
Property  & Minimum Value & Maximum Value \\
\hline
$M_r-5\log(h)$      &  $-22$   & $-18$  \\ 
$(g-r)$             &  $0.2$    & $0.9$   \\
$\mu_r$             &  $19$    & $23$   \\
$eclass$       &  $-0.2$    & $0.4$    \\
$C=r_{90}/r_{50}$   &  $1.8$    & $3.4$   \\ 
\hline
\label{cutoffs}
\end{tabular}
\end{table}

\subsection{Galaxy parameters}
The SDSS provides several photometric and spectroscopic 
parameters of the surveyed galaxies. Among the available data for each object in DR4,
we have used in our analyses parameters that are related to different 
physical properties of the galaxies: luminosity, star formation rate, 
light distribution inside the galaxies and the dominant stellar populations.  
The galaxy parameters we have focused our study on are: 
\begin{enumerate}
\item $r-$band absolute magnitude, $M_r$.
\item $g-r$ colour.
\item The mono-parametric spectral classification based on the eigentemplates
expansion of galaxy's spectrum $eclass={\rm atan}
(-ecoeff_2/ecoeff_1)$. This parameter ranges from about $-0.35$ 
for early-type galaxies to $0.5$ for late-type galaxies.
The galaxy spectral classification eigentemplates were created from a sample 
of about 200,000 spectra. The eigenspectra are an early version of those created 
by \citet{yip04}. 
\item $r-$band surface brightness, $\mu_r$, computed inside the radius that
encloses 50\% of Petrosian flux, $r_{50}$.
\item  $r-$band concentration parameter defined as the ratio
between the radii that enclose 90\%  and 50\% of the Petrosian flux,
$C=r_{90}/r_{50}$. Typically, early-type galaxies have $C>2.5$, while
for late-types $C<2.5$ \citep{st01}. 
\end{enumerate}

\begin{figure}
\includegraphics[width=90mm]{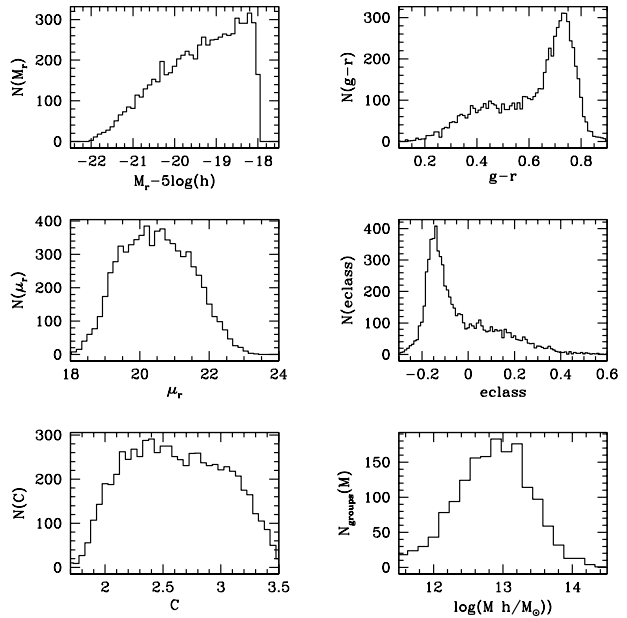}
\caption{
The distributions of galaxy properties in our sample. Lower right panel
shows the $\log(M)$ distribution for our sample of groups.
}
\label{fig1}
\end{figure}

We list in Table \ref{cutoffs} the parameters cut-offs we have adopted
for the present analyses, and in Figure \ref{fig1} we show the corresponding 
distributions for the galaxies in our sample. 

Galaxy magnitudes used throughout this paper have been corrected for Galactic extinction 
following \citet{sch98}, absolute magnitudes have been computed assuming 
$\Omega_0=0.3$, $\Omega_{\Lambda}=0.7$ and $H_0=100~h~{\rm km~s^{-1}~Mpc^{-1}}$ 
and $K-$corrected using the method of \citet{blanton03}~({\small KCORRECT} version 4.1). 
All magnitudes are in the AB system.
\section{Dependence of galaxy properties on group mass}
In order to determine which galaxy properties are more correlated with the mass of the 
group where the galaxy is located, we have followed here the \citet{blanton05} 
approach.  Firstly, we consider the variance of the logarithm of the galaxy's parent 
group mass, 
$\log(M)$:
\begin{equation}
\sigma^2=\frac{1}{n-1}\sum_{i=1}^{n}(\log(M_i)-\overline{\log(M)})^2 ,
\label{sigma}
\end{equation}
where  $\overline{\log(M)}$ is the mean value of $\log(M)$ for the $n$ galaxies in 
the sample. Then, we measure the mean value of $\log(M)$ as a function of a 
given galaxy property $X$. For doing this, we split galaxies into $m$ bins 
centred in the values $X_j$ ($j=1,m$), and compute:
\begin{equation}
\overline{\log(M)}_j=\frac{1}{N_j}\sum_{|X_i-X_j|\leq\Delta X/2}\log(M_i)
\end{equation}
where $N_j$ is the number of galaxies in the $j$ bin which is $\Delta X$ wide.
We finally compute the quantity defined as:
\begin{equation}
\sigma^2_X=\frac{1}{n-1}\sum_{j=1}^m 
\sum_{|X_i-X_j|\leq\Delta X/2}(\log(M_i)-\overline{\log(M)}_j)^2 \leq\sigma^2.
\label{sigmaX}
\end{equation}
Hence, $\sigma_X^2$ represents the variance of $\log(M)$ after subtracting the 
global trends of $\log(M)$ with the parameter $X$ and so the property most 
closely related to group mass will minimise $\sigma_X$.
This can be straightforwardly generalised to two properties $X$ and 
$Y$ if one wants to analyse which pair of properties are most closely 
correlated with mass. 
The quantity $\sigma_X$ is independent of the units of the physical quantity $X$, 
but it does depend on the choice of binning for it.
We have taken care of this by ensuring that each bin 
is bigger than the mean errors in the considered parameter, is smaller than
the features in the parameter's distribution 
and contains a large enough number of galaxies.

\begin{figure}
\includegraphics[width=90mm]{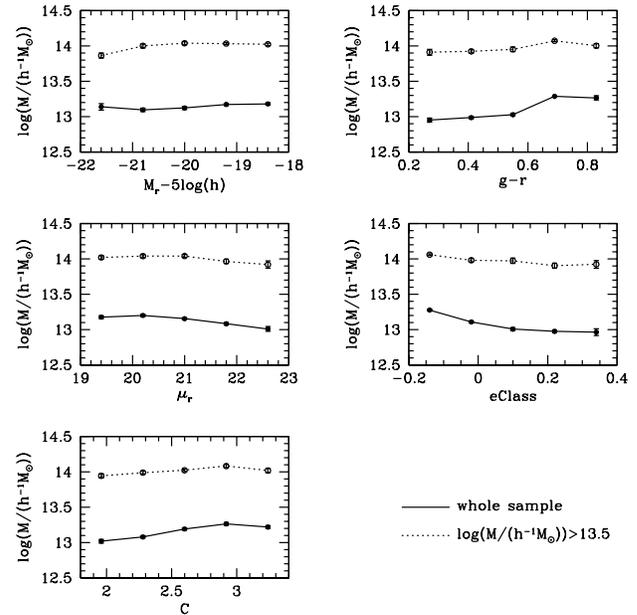}
\caption{
Mean value of $\log(M)$ as a function of galaxy properties.
Continuous lines are the results for the whole sample of groups used
in this work, while dotted lines correspond to the 
$\log(M/(h^{-1}M_{\odot}))\ge 13.5$ subsample. 
Error bars were estimated using the bootstrap resampling technique.
}
\label{MX}
\end{figure}

In Table \ref{tab1} we list the values of the differences $\sigma_X^2-\sigma^2$
and $\sigma_{XY}^2-\sigma^2$ for our sample of galaxies in groups. 
The single most predictive quantity for group
mass is $g-r$ colour, in the second place appears the spectral parameter 
$eclass$ and in third place comes the concentration parameter $C$.
We show in Figure \ref{MX} in solid lines the mean value of $\log(M)$ as
a function of the 5 galaxy properties. 
The most predictive pair of properties is $g-r$ colour
and absolute magnitude taken together, closely followed by the pairs 
$eclass/M_r$, $g-r/\mu_r$ and $C/\mu_r$. 

\citet{yo02} found that groups more massive than $M\sim 10^{13.5}M_{\odot}$
have galaxy populations that differ significantly from the typical 
field population.  In these groups, the fraction of low and non-star forming 
galaxies is higher than in the field and this difference increases with mass.
Consistently, \citet{mardom02} found that spectral type segregation
is present in these more massive groups. With these results in mind,
we have repeated our analysis to a subsample of galaxies in groups 
with masses $M>10^{13.5}h^{-1}M_{\odot}$. In this high mass subsample we have
1698 galaxies in 320 groups. 
The results are shown in Table \ref{tab2} and the mean value
of $\log(M)$ as a function of the single properties are shown in dotted lines in
Figure \ref{MX}.
The only change with respect to the whole sample of groups
is in the pairs, the most predictive pair is now $eclass/M_r$ and the
second one is $g-r/M_r$, but the differences in their $\sigma_X$ are small.

A special care must be taken regarding the possible presence of brightest
group galaxies that may be by far the principal contributor to the group's 
total luminosity. This could bias the results for high mass groups where
the brightest group galaxies are expected to be particularly bright.  
Therefore, we have re-done the calculations excluding the brightest
object in each group and found that the results hold.

We performed a number of tests to analyse the reliability of our
results: restricting the samples by the number of galaxy members in the groups;
splitting the samples into 2 roughly equal number subsamples according to 
different criteria such as the region in the sky according to their right ascension and 
into those that have an even (odd) identification number in the 
\citet{zmm06} catalogue. The results of these tests are summarised in Table \ref{tests}. 
We observe that the ranking of single properties is stable, it is the same for
all but for the subsample of high mass groups with at least 8 members, where
there is an inversion between the second and the third most predictive properties.
In all cases the single most predictive quantity is the $g-r$ colour.
Regarding the pairs of properties, most subsamples give the same first pair:
colour with absolute magnitude. The exceptions are the two high mass subsamples.
It is clear that the second most predictive pair is not stable.

\begin{table}
\caption{Galaxy parameters as group mass indicators.
The quantity $\sigma^2$ is variance of $\log(M)$, 
$\sigma_X^2$ is the variance around the
mean values of $\log(M)$ for each parameter $X$ and $\sigma_{XY}^2$ is the 
variance around the mean relation for the pairs or properties $X$ and $Y$. 
Quoted values are expressed in units of $10^{-3}$.}
\begin{tabular}{lcccccc}
\hline
 & & \multicolumn{5}{c}{Property Y}\\
Property X & $\sigma_X^2-\sigma^2$ & $M_r$ & $g-r$ & $\mu_r$ & 
$eclass$ & $C$ \\
 & & \multicolumn{5}{c}{$\sigma_{XY}^2-\sigma^2$}  \\
\hline
$M_r$    & $-1 $        & ...          & \fbox{$-27$} & $-6 $ & $-25$ & $-12$ \\
$g-r$    & \fbox{$-20$} & \fbox{$-27$} & ...          & $-25$ & $-23$ & $-22$ \\
$\mu_r$  & $-3 $        & $-6 $        & $-25$        & ...   & $-19$ & $-25$ \\
$eclass$ & $-16$        & $-25$        & $-23$        & $-19$ & ...   & $-19$ \\
$C$      & $-8 $        & $-12$        & $-22$        & $-25$ & $-19$ & ...   \\
\hline
\label{tab1}
\end{tabular}
\end{table}

\begin{table}
\caption{Similar to Table \ref{tab1} for massive
($\log(M/(h^{-1}M_{\odot}))>13.5$) groups only.  Quoted values are expressed 
in units of $10^{-4}$.}
\begin{tabular}{lcccccc}
\hline
 & & \multicolumn{5}{c}{Property Y}\\
Property X & $\sigma_X^2-\sigma^2$ & $M_r$ & $g-r$ & $\mu_r$ & 
$eclass$ & $C$ \\
 & & \multicolumn{5}{c}{$\sigma_{XY}^2-\sigma^2$}  \\
\hline
$M_r$    & $-9 $        & ...          & $-59$ & $-33$ & \fbox{$-60$} & $-39$ \\
$g-r$    & \fbox{$-40$} & $-59$        & ...   & $-50$ & $-51$        & $-56$ \\
$\mu_r$  & $-9 $        & $-33$        & $-50$ & ...   & $-37$        & $-30$ \\
$eclass$ & $-26$        & \fbox{$-60$} & $-51$ & $-37$ & ...          & $-43$ \\
$C$      & $-17$        & $-39$        & $-56$ & $-30$ & $-43$        & ...   \\
\hline
\label{tab2}
\end{tabular}
\end{table}

\begin{table}
\caption{Galaxy parameters as group mass indicators: tests of stability.
First column indicates the subsample (see text), second column lists
the ranking of the most predictive single properties, last column
lists the first two most predictive pairs of properties.
}
\begin{tabular}{lcc}
\hline
\multicolumn{3}{c}{Groups with $\geq4$ members}\\
\hline
Subsample   & Single parameter ranking & First two pairs\\
\hline
All groups       & $g-r, eclass, C, \mu_r, M_r$ & $g-r/M_r, g-r/\mu_r$  \\
High mass        & $g-r, eclass, C, \mu_r, M_r$ & $eclass/M_r, g-r/M_r$ \\
$\alpha<12^h$    & $g-r, eclass, C, \mu_r, M_r$ & $g-r/M_r, g-r/\mu_r$  \\
$\alpha\geq12^h$ & $g-r, eclass, C, \mu_r, M_r$ & $g-r/M_r, M_r/eclass$  \\
even             & $g-r, eclass, C, \mu_r, M_r$ & $g-r/M_r, M_r/eclass$  \\
odd              & $g-r, eclass, C, \mu_r, M_r$ & $g-r/M_r, M_r/eclass$  \\
\hline
\multicolumn{3}{c}{Groups with $\geq8$ members}\\
\hline
All groups       & $g-r, eclass, C, \mu_r, M_r$ & $g-r/M_r, g-r/eclass$  \\
High mass        & $g-r, C, eclass, \mu_r, M_r$ & $g-r/C, g-r/eclass$ \\
$\alpha<12^h$    & $g-r, eclass, C, \mu_r, M_r$ & $g-r/M_r, g-r/C$  \\
$\alpha\geq12^h$ & $g-r, eclass, C, \mu_r, M_r$ & $g-r/M_r, g-r/eclass$  \\
even             & $g-r, eclass, C, \mu_r, M_r$ & $g-r/M_r, g-r/eclass$  \\
odd              & $g-r, eclass, C, \mu_r, M_r$ & $g-r/M_r, g-r/eclass$  \\
\hline
\label{tests}
\end{tabular}
\end{table}
\section{Dependence of Galaxy properties on group-centric distances}
As said before, \citet{mardom02} found a trend of the fraction of 
different galaxy types with the distance to the centre of the groups,
with low and non-star forming galaxies being located preferentially
towards the centres. Therefore, similar behaviours are to be expected
for other galaxy parameters such as colour or concentration index.
In a similar way as we have done in the previous section, we study here
which galaxy property, or pair of properties, is most closely related to galaxy
distance from the group centre. We have applied here the same statistics as in the
previous section, replacing the logarithm of the
group mass in equations \ref{sigma} to \ref{sigmaX} by the group centric 
distance in units of the group virial radius, $r_{\rm vir}$, 
computed by \citet{zmm06}.
We have restricted the sample of groups to those that have at least 8 members, to
ensure ourselves that determinations of group centre positions are more robust,
with this restriction we have 2662 galaxies in 428 groups.

\begin{figure}
\includegraphics[width=90mm]{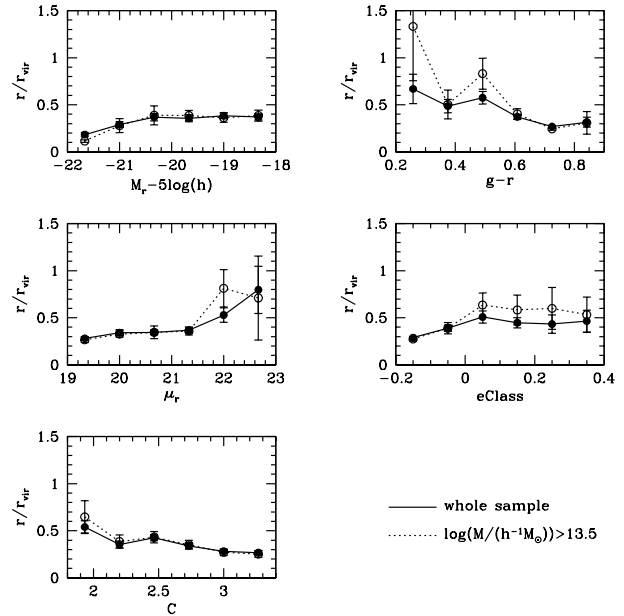}
\caption{
Mean value of $r/r_{\rm vir}$ as a function of galaxy properties.
Continuous lines are the results for the whole sample of groups used
in this work, while dotted lines correspond to the 
$\log(M/(h^{-1}M_{\odot}))\ge 13.5$ subsample. 
}
\label{RX}
\end{figure}

In Figure \ref{RX} we show the mean values of $r/r_{\rm vir}$ as a function of
the parameters for all groups with at least 8 members and for the high mass
subsample.
The resulting values for $\sigma_X^2-\sigma^2$ and $\sigma_{XY}^2-\sigma^2$ 
are quoted in Table \ref{tab3}. The $g-r$ colour is the single parameter
that correlates best with the distance from the group centre, 
the second place corresponds to surface brightness. 
Among the pairs of parameters, in the first place is $g-r/eclass$. 
When analysing the high mass subsample which comprises 1387 galaxies in
166 groups (Table \ref{tab4} and
Figure \ref{RX}), we find the same ranking for the single parameters
and the same first pair, but a different order in the remaining pairs.

We have performed similar tests as in previous section to evaluate the stability
of the results. This is summarised in Table \ref{testsR}.
We find that colour is the parameter most correlated with group-centric distance
in all subsamples, and in most of them, the surface brightness is the following one.
The differences between the single parameter ranking for the different subsamples
are among the second to the fourth places that are taken by the quantities
$\mu_r$, $eclass$ and $C$. But as can be seen in Table \ref{tab3}, their values 
$\sigma_X^2-\sigma^2$ do not differ significantly.
For the pairs of properties, in most cases the first place corresponds to colour and
spectral type, while the second ranked pair is not stable at all.

\begin{table}
\caption{Galaxy parameters as group-centric distance indicators.
The quantity $\sigma^2$ is variance of $r/r_{\rm vir}$, 
$\sigma_X^2$ is the variance around the
mean values of $r/r_{\rm vir}$ for each parameter $X$ and $\sigma_{XY}^2$ is the 
variance around the mean relation for the pairs or properties $X$ and $Y$. 
Quoted values are expressed in units of $10^{-3}$.}
\begin{tabular}{lcccccc}
\hline
 & & \multicolumn{5}{c}{Property Y}\\
Property X & $\sigma_X^2-\sigma^2$ & $M_r$ & $g-r$ & $\mu_r$ & $eclass$ & $C$ \\
 & & \multicolumn{5}{c}{$\sigma_{XY}^2-\sigma^2$}  \\
\hline
$M_r$    & $-1 $        & ...   & $-21$        & $-8 $ & $-11$        & $-7$  \\
$g-r$    & \fbox{$-14$} & $-21$ & ...          & $-23$ & \fbox{$-59$} & $-25$ \\
$\mu_r$  & $-7 $        & $-8 $ & $-23$        & ...   & $-53$        & $-24$ \\
$eclass$ & $-6 $        & $-11$ & \fbox{$-59$} & $-53$ & ...          & $-11$ \\
$C$      & $-6 $        & $-7$  & $-25$        & $-24$ & $-11$        & ...   \\
\hline
\label{tab3}
\end{tabular}
\end{table}

\begin{table}
\caption{Similar to Table \ref{tab3} for massive
($\log(M/(h^{-1}M_{\odot}))>13.5$) groups only.  Quoted values are expressed 
in units of $10^{-3}$.}
\begin{tabular}{lcccccc}
\hline
 & & \multicolumn{5}{c}{Property Y}\\
Property X & $\sigma_X^2-\sigma^2$ & $M_r$ & $g-r$ & $\mu_r$ & $eclass$ & $C$ \\
 & & \multicolumn{5}{c}{$\sigma_{XY}^2-\sigma^2$}  \\
\hline
$M_r$    & $-2 $        & ...    & $-118$        & $-57$  & $- 3$         & $-21$  \\
$g-r$    & \fbox{$-47$} & $-118$ & ...           & $-64$  & \fbox{$-304$} & $-242$ \\
$\mu_r$  & $-19$        & $-57$  & $-64$         & ...    & $-58$         & $-151$ \\
$eclass$ & $-17$        & $-3$   & \fbox{$-304$} & $-58$  & ...           & $-163$ \\
$C$      & $-9 $        & $-21$  & $-242$        & $-151$ & $-163$        & ...    \\
\hline
\label{tab4}
\end{tabular}
\end{table}

\begin{table}
\caption{Galaxy parameters as group-centric distance indicators: tests of stability.
First column indicates the subsample (see text), second column lists
the ranking of the most predictive single properties, last column
lists the first two most predictive pairs of properties.
}
\begin{tabular}{lcc}
\hline
Subsample   & Single parameter ranking & First two pairs\\
\hline
All groups       & $g-r, \mu_r, eclass, C, M_r$ & $g-r/eclass, eclass/\mu_r$ \\
High mass        & $g-r, \mu_r, eclass, C, M_r$ & $g-r/eclass, g-r/C$        \\
$\alpha<12^h$    & $g-r, \mu_r, eclass, C, M_r$ & $g-r/eclass, g-r/C$        \\
$\alpha\geq12^h$ & $g-r, C, eclass, \mu_r, M_r$ & $g-r/\mu_r, g-r/C$         \\
even             & $g-r, \mu_r, eclass, C, M_r$ & $g-r/eclass, g-r/M_r$      \\
odd              & $g-r, eclass, C, \mu_r, M_r$ & $g-r/eclass, g-r/C$        \\
\hline
\label{testsR}
\end{tabular}
\end{table}
\section{Discussion and conclusions}
By using a large sample of galaxies in groups in the SDSS DR4, we have analysed
how the properties of galaxies in groups are related to the environment. 
We find that the $g-r$ colour is the parameter most dependent on group mass.
The pair of properties that is most correlated to group mass are
$g-r$ colour and $r-$band absolute magnitude taken together. 
Our results do not change when we consider massive groups ($M>10^{13.5}h^{-1}M_{\odot}$). 
Regarding the position inside the systems, again we find that colour is the most 
predictive property for group-centric distances, while colour and $eclass$
comprise the most predictive pair. The results do not vary when we 
exclude low mass systems.

It should be noticed that both, the single and the pair of most predictive 
galaxy properties as a function of the group environment, are nearly the same
ones that \citet{blanton05} found for field galaxies using the local density. The 
only difference appears 
in the pair of properties when the group-centric distance is considered. 
\citet{blanton05} found that colour and luminosity are the most predictive 
pair of properties, while according to our results, the pair colour/$eclass$ 
is the most relevant.  
The similarity between our results for galaxies in groups and those 
by \citet{blanton05} for field galaxies suggests that the physical process 
associated with the galaxy formation and evolution are similar for these 
two environments. 

It should be noted that the study of high mass systems gives similar results 
to those of the whole sample. This might be indicating that, over an
important range of environments, galaxies are affected in a similar way.    
From field to high mass groups it is expected the merger rate to vary,
nevertheless, it is remarkable that the same galaxy property is the one 
that correlate best with the environment. It would be interesting to repeat
the present analysis for massive clusters of galaxies.
  
It is also interesting to note that the 
concentration parameter, closely related to the galaxy morphology, is 
not good in predicting the group environment. This is particularly surprising 
considering the well known effect of morphological segregation in systems of
galaxies. This result indicates that the morphological transformations that
produce these relations are in a sense a byproduct of the transformations that 
produce the environmental trends on other properties like colour.
\section*{Acknowledgements}
We thank the anonymous referee for helpful comments that improved this paper. 
This work has been partially supported with grants from Consejo Nacional
de Investigaciones Cient\'\i ficas y T\'ecnicas de la Rep\'ublica Argentina
(CONICET), Secretar\'\i a de Ciencia y Tecnolog\'\i a de la Universidad 
de C\'ordoba and Agencia Nacional de Promoci\'on Cient\'\i fica y
Tecnol\'ogica, Argentina.

Funding for the Sloan Digital Sky Survey (SDSS) has been provided by the 
Alfred P. Sloan 
Foundation, the Participating Institutions, the National Aeronautics and Space 
Administration, the National Science Foundation, the U.S. Department of Energy, 
the Japanese Monbukagakusho, and the Max Planck Society. The SDSS Web site is 
http://www.sdss.org/.
The SDSS is managed by the Astrophysical Research Consortium (ARC) for the 
Participating Institutions. The Participating Institutions are The University 
of Chicago, Fermilab, the Institute for Advanced Study, the Japan Participation 
Group, The Johns Hopkins University, the Korean Scientist Group, Los Alamos 
National Laboratory, the Max Planck Institut f\"ur Astronomie (MPIA), the 
Max Planck Institut f\"ur Astrophysik (MPA), New Mexico State University, 
University of Pittsburgh, University of Portsmouth, Princeton University, 
the United States Naval Observatory, and the University of Washington.

\label{lastpage}

\begin{thebibliography}{}
\bibitem[Adelman-McCarthy et al.(2006)]{dr4}
Adelman-McCarthy J.~K., et al.\ 2006, \apjs, 162, 38 
\bibitem[\protect\citeauthoryear{Blanton et al.}{2003}]{blanton03} 
Blanton, M.~R., et al.\ 2003, \aj, 125, 2348 
\bibitem[Blanton et al.(2005)]{blanton05} Blanton, M.~R., 
Eisenstein, D., Hogg, D.~W., Schlegel, D.~J., \& Brinkmann, J.\ 2005, \apj, 
629, 143 
\bibitem[\protect\citeauthoryear{Dom{\'{\i}}nguez et al.}{2002}]{mardom02} 
Dom{\'{\i}}nguez, M.~J., Zandivarez, A.~A., Mart{\'{\i}}nez, H.~J., 
Merch{\' a}n, M.~E., Muriel, H., \& Lambas, D.~G.\ 2002, \mnras, 335, 825 
\bibitem[Dressler(1980)]{dress80} Dressler, A.\ 1980, \apj, 
236, 351 
\bibitem[G{\'o}mez et al.(2003)]{gomez} G{\'o}mez, P.~L., et 
al.\ 2003, \apj, 584, 210 
\bibitem[\protect\citeauthoryear{Hashimoto \& Oemler}{2000}]{ho00} 
Hashimoto, Y., \& Oemler, A.~J.\ 2000, \apj, 530, 652
\bibitem[\protect\citeauthoryear{Mart{\'{\i}}nez et al.}{2002}]{yo02} 
Mart{\'{\i}}nez, H.~J., Zandivarez, A., Dom{\'{\i}}nguez, M., Merch{\' a}n, 
M.~E., \& Lambas, D.~G.\ 2002, \mnras, 333, L31 
\bibitem[Merch{\'a}n \& Zandivarez(2005)]{mz05} Merch{\'a}n, 
M.~E., \& Zandivarez, A.\ 2005, \apj, 630, 759 
\bibitem[\protect\citeauthoryear{Schlegel et al.}{1998}]{sch98} Schlegel, D.~J., 
Finkbeiner, D.~P., \& Davis, M.\ 1998, \apj, 500, 525 
\bibitem[\protect\citeauthoryear{Strateva et al.}{2001}]{st01} 
Strateva, I., et al.\ 2001, \aj, 122, 1861 
\bibitem[Strauss et al.(2002)]{mgs}
Strauss, M.~A., et al.\ 2002, \aj, 124, 1810
\bibitem[\protect\citeauthoryear{Yip et al.}{2004}]{yip04} Yip, C.~W., et al.\ 2004, 
\aj, 128, 585 
\bibitem[\protect\citeauthoryear{York et al.}{2000}]{york00} 
York, D.~G., et al.\ 2000, \aj, 120, 1579 
\bibitem[\protect\citeauthoryear{Zabludoff \& Mulchaey}{1998}]{zm98} 
Zabludoff, A.~I., \& Mulchaey, J.~S.\ 1998, \apj, 496, 39 
\bibitem[\protect\citeauthoryear{Zabludoff \& Mulchaey}{2000}]{zm00} 
Zabludoff, A.~I., \& Mulchaey, J.~S.\ 2000, \apj, 539, 136 
\bibitem[Zandivarez et al.(2006)]{zmm06}
Zandivarez, A., Mart\'\i nez, H.~J., \& Merch\'an, M.~E.\ 2006, \apj, in press, 
preprint (astro-ph/0602405)

\end{thebibliography}
\end{document}